\documentclass[twocolumn,prl,longbibliography]{revtex4-1}

\usepackage{graphicx}
\usepackage{amsmath}
\usepackage{amsfonts}
\usepackage{color}
\usepackage{centernot}
\usepackage{mathtools}
\usepackage{stmaryrd}

\begin{document}

\title{Hidden scale invariance of intermittent turbulence in a shell model}
\author{Alexei A. Mailybaev}\email{alexei@impa.br}
\affiliation{Instituto de Matem\'atica Pura e Aplicada -- IMPA, 22460-320 Rio de Janeiro, Brazil}

\begin{abstract}
It is known that scale invariance is broken in the developed hydrodynamic turbulence due to intermittency, substantiating complexity of turbulent flows. Here we challenge the concept of broken scale invariance by establishing a hidden self-similarity in intermittent turbulence. Using a simplified (shell) model, we derive a nonlinear spatiotemporal scaling symmetry of inviscid equations, which are reformulated in terms of intrinsic times introduced at different scales of motion. Numerical analysis persuasively confirms 
that this symmetry is restored in a statistical sense within the inertial interval. At the end, we discuss implications of this result for the Navier--Stokes system.
\end{abstract}

\maketitle

Exact or approximate scaling symmetries allow accessing non-trivial properties of complex physical systems through self-similarity and renormalization~\cite{barenblatt1996scaling,cardy1996scaling}. In hydrodynamic turbulence, scale invariance is considered in a statistical sense, i.e., satisfied by a stationary probability distribution~\cite{frisch1999turbulence,falkovich2009symmetries}. Kolmogorov's theory of 1941~\cite{kolmogorov1941local}, which describes the homogeneous and isotropic developed turbulence, assumes that the statistically stationary state is scale invariant in the inertial interval. Such interval comprises a wide range of scales, which separates the forcing range at large scales from the dissipative range at very small scales. Viscosity can be neglected in the inertial interval, which yields the scaling symmetry for velocity fields $\mathbf{v}(\mathbf{r},t)$ in the form
	\begin{equation}
	\label{eqSM0}
	t,\mathbf{r},\mathbf{v} \mapsto \alpha^{1-h}t,\alpha\mathbf{r},\alpha^h\mathbf{v},
	\end{equation}
with an arbitrary exponent $h \in \mathbb{R}$ and $\alpha > 0$. Then, the exponent $h = 1/3$ is fixed by the existence of energy cascade. However, in real turbulent flows this scale invariance is broken by the intermittency phenomenon: the flow displays activity only within a fraction of time or space, which decreases with the scale under consideration~\cite{frisch1999turbulence}. This broken symmetry leads to a much higher complexity of turbulent flows, whose full theoretical description remains an open problem. 

Here we argue that breaking of scale invariance by intermittency is illusive, by demonstrating that self-similarity emerges with respect to a hidden nonlinear scaling symmetry. This new symmetry is established for equations of inviscid dynamics written in terms of intrinsic times introduced at different scales of motion. We explicitly derive this symmetry for a shell model of turbulence~\cite{l1998improved}, and verify the asserted self-similarity numerically in fine detail. This model successfully reproduces the fundamental properties of turbulence such as the energy cascade with inertial interval and intermittency with anomalous scaling exponents~\cite{biferale2003shell}, therefore, providing a proper testing platform for our ideas. The proposed construction is inspired by the universality of Kolmogorov multipliers~\cite{kolmogorov1962refinement}, defined by ratios of velocity increments at adjacent scales, which has been confirmed for shell models~\cite{benzi1993intermittency,eyink2003gibbsian,biferale2017optimal}, and then observed in direct numerical simulations and experimental data~\cite{chen2003kolmogorov}. Our rescaled variables are analogous to Kolmogorov's multipliers, but it is the hierarchy of intrinsic times at different scales that turns such kind of universality into the exact scaling symmetry. 

\textbf{Scale invariance broken by intermittency.} Shell models mimic the multi-scale structure of turbulence by means of discrete spatial scales (shells) with the wavenumbers $k_n = k_0\lambda^n$, where  $n = 1,2,\ldots$ and the inter-shell ratio is $\lambda = 2$~\cite{gledzer1973system,ohkitani1989temporal,biferale2003shell} . In the Sabra model~\cite{l1998improved}, each shell is characterized by a complex shell velocity $u_n(t)$ depending on time and modeling velocity fluctuations at the corresponding scale. Equations of motion are constructed to mimic the structure of incompressible three-dimensional Navier--Stokes equations in the form
	\begin{equation}
	\label{eqSM}
	\begin{array}{rcl}
	\displaystyle
	\frac{du_n}{dt} & = & 
	\displaystyle
	i\Big(k_{n+1}u_{n+2}u_{n+1}^*
	-\frac{1}{2}k_nu_{n+1}u_{n-1}^* \\[7pt]
	&& 
	\displaystyle
	+\,\frac{1}{2}k_{n-1}u_{n-1}u_{n-2}\Big)-\nu k_n^2u_n+f_n,
	\end{array}
	\end{equation}
where $\nu$ is the viscosity and the forcing terms $f_n$ are limited to large scales, i.e., to a few initial shells with $n \sim 1$. This system has two inviscid invariants corresponding to energy $E = \sum |u_n|^2$ and helicity $H = \sum (-1)^nk_n |u_n|^2$.  Considering the forcing with amplitude $|f_n| \sim F$, we define the integral length $L = 1/k_0$, the integral velocity $U = \sqrt{F/k_0}$, and the Reynolds number $\mathrm{Re} = UL/\nu$. 

\begin{figure}
\centering
\includegraphics[width=0.48\textwidth]{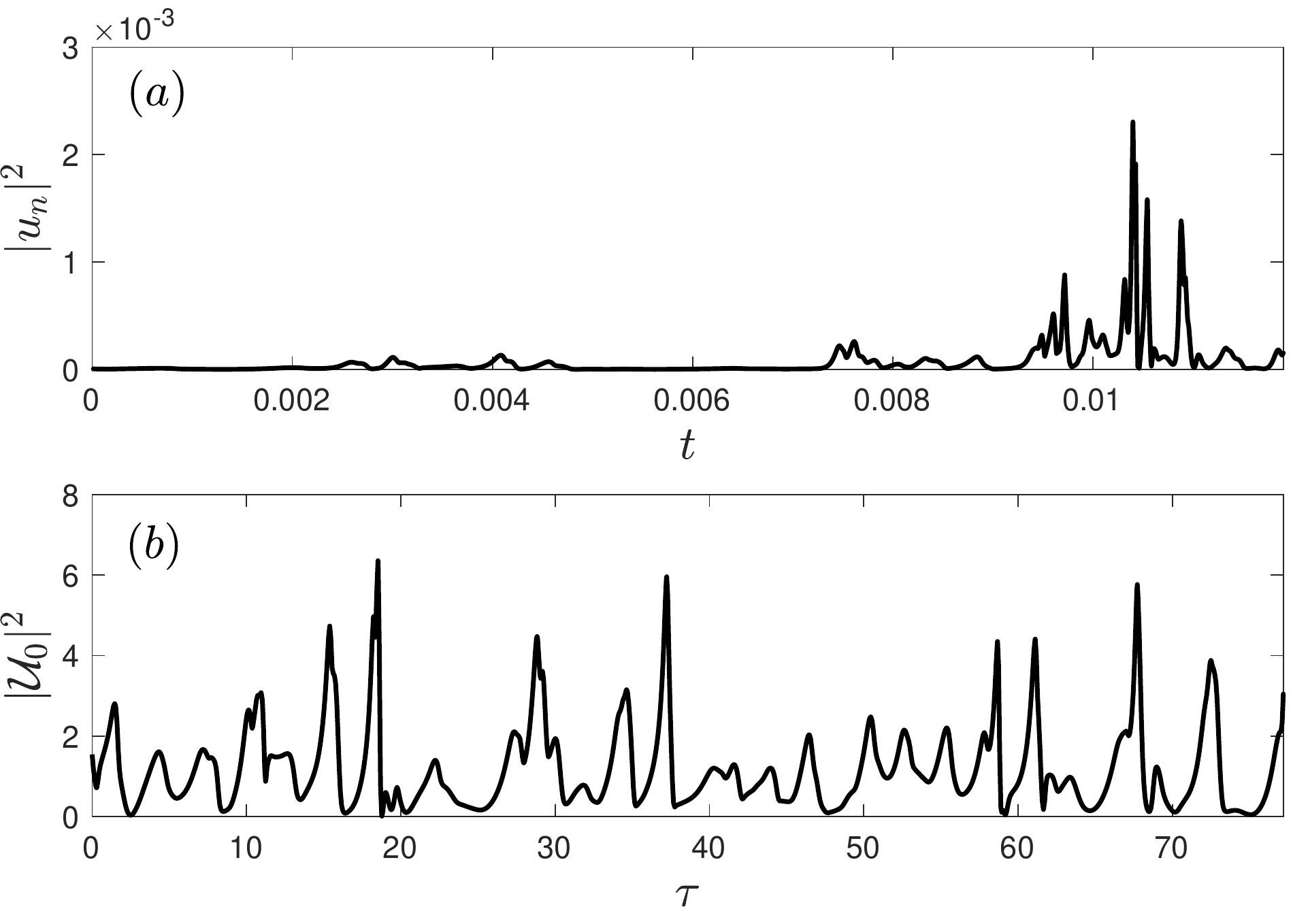}
\caption{(a) Typical evolution of the squared amplitude for the shell velocity in the inertial interval, $|u_n|^2$ with $n = 20$. (b) Evolution of the respective rescaled variable, $|\mathcal{U}_0(\tau)|^2 = k_m^2T_m^2|u_m|^2$ for the same reference shell $m = 20$. Intermittent fluctuations are regularized as a result of dynamical temporal scaling. For details of the simulation, see later in the text.}
\label{fig1}
\end{figure}

For very large Reynolds numbers, developed turbulence in Sabra model features a large inertial interval of scales (shells $n$) separating the forcing range at large scales (small $n$) from the dissipative range at small scales (large $n$). In the inertial range, both viscous and forcing terms  can be neglected, which yields the inviscid unforced system by setting $\nu = 0$ and $f_n = 0$ in (\ref{eqSM}). The scale invariance of this system, analogous to (\ref{eqSM0}), is given by the transformations $t,u_n \mapsto \alpha^{1-h}t,\alpha^hu_{n+s}$, where $h \in \mathbb{R}$ and $\alpha = \lambda^{-s}$ takes discrete values with arbitrary integer $s$ such that $k_n/\alpha = k_{n+s}$. According to the multi-fractal theory~\cite{frisch1999turbulence}, the intermittent  turbulence in Sabra model develops motions with a continuous range of exponents $h$, therefore, breaking scale invariance in the inertial interval. Such intermittency manifests itself in temporal fluctuations of velocities: short streaks of fast large-amplitude oscillations are separated by long periods of slow motion, becoming more and more pronounced at smaller scales; see Fig.~\ref{fig1}(a). The intermittency is quantified by the scaling of time-averaged velocity moments, $\langle|u_n|^p\rangle \propto k_n^{-\zeta_p}$, with anomalous exponents $\zeta_p$ depending nonlinearly on $p$ in close agreement with structure functions in hydrodynamic turbulence~\cite{l1998improved}.

\textbf{Intrinsic times and rescaled velocities.} The central idea is to cope with the intermittency by adjusting temporal scales dynamically, according to the momentary intensity of motion at every scale. Feasibility of such dynamical scaling relies on its combination with the translational temporal symmetry, such that different scaling laws are introduced locally at different times and scales. As we shall see, this results in nonlinear and time-dependent relations defining a new form of exact scaling symmetry for the inviscid equations.

Let us consider some \textit{reference shell} $m$ within the inertial interval. We introduce the momentary temporal scale associated to this shell as
	\begin{equation}
	\label{eqS1}
	T_m(t) = \Big(k_0^2U^2+\sum_{n < m} k_n^2|u_n(t)|^2\Big)^{-1/2}.
	\end{equation}
Here the expression in parentheses can be interpreted as an enstrophy contained in shells $n < m$, and the term $k_0^2U^2$ characteristic for the forcing range ensures that $T_m$ is bounded. In the inertial interval, the time-averaged structure function $\langle |u_n|^2 \rangle \propto k_n^{-\zeta_2}$ has the anomalous scaling exponent $\zeta_2 \approx 0.72$~\cite{l1998improved}. Therefore, the sum in (\ref{eqS1}) converges as a geometric progression with the dominant contribution from shells close to $m$, justifying our choice of $T_m$ as the local temporal scale at shell $m$ and time $t$. 

We now introduce the dimensionless time and shell velocity, whose scales are adjusted dynamically as
	\begin{equation}
	\label{eqS2}
	\tau = \int_0^t \frac{dt'}{T_m(t')}, \quad
	\mathcal{U}_N = k_{m}T_m(t) u_{N+m}(t),
	\end{equation}
with integer indices $N > -m$.
Note that the new time $\tau$ is a nonlinear solution-dependent function of physical time $t$, designed such that its increments $d\tau = dt/T_m$ are calibrated with the momentary temporal scale $T_m(t)$. We call $\tau$ the \textit{intrinsic time} at shell $m$, and consider relations (\ref{eqS2}) as implicit definition of the rescaled shell velocity $\mathcal{U}_N(\tau)$. A numerical example of this new function is presented in Fig.~\ref{fig1}(b), obtained from the original variables in Fig.~\ref{fig1}(a). Since the dynamical temporal scale $T_m$ follows local intermittent changes at shell $m$, the resulting rescaled velocity displays regular fluctuations with typical periods and amplitudes of order unity. One can draw analogy of $\mathcal{U}_0$ from (\ref{eqS2}) with a  Kolmogorov multiplier $u_m/u_{m-1}$~\cite{benzi1993intermittency,eyink2003gibbsian}, because $k_mT_m \sim |u_{m-1}|^{-1}$.

Rescaled variables $\mathcal{U}_N(\tau)$ are uniquely expressed by (\ref{eqS2}) in terms of original velocities $u_n(t)$. In this paragraph we verify that the converse is also true. Let us assume that all variables $\mathcal{U}_N(\tau)$ are known for some reference shell $m$. Then, we derive rescaled variables (\ref{eqS2})  corresponding to the next reference shell $m^+ = m+1$ in the form
	\begin{equation}
	\label{eqS2b_1}
	\tau^{+}= \int_0^{\tau} \!\sqrt{1+\big|\mathcal{U}_0\big|^2} d\tau,\quad
	\mathcal{U}_N^{+} 
	= \frac{2 \,\mathcal{U}_{N+1}}{\sqrt{1+|\mathcal{U}_0|^2}},
	\end{equation}
as one can check using (\ref{eqS1}). Similarly, the rescaled variables corresponding to the previous reference shell $m^- = m-1$ are found as
	\begin{equation}
	\label{eqS2b_2}
	\tau^{-} = \int_0^{\tau} \!\sqrt{1-\frac{|\mathcal{U}_{-1}|^2}{4}} d\tau,\quad
	\mathcal{U}_N^{-} 
	= \frac{\mathcal{U}_{N-1}}{\sqrt{4-|\mathcal{U}_{-1}|^2}}.
	\end{equation}
These formulas can be used iteratively to define one-to-one relations between representations (\ref{eqS2}) introduced for any pair of reference shells $m$ and $m'$. In particular, for $m' = 1$, expressions (\ref{eqS1})--(\ref{eqS2}) yield $\tau' = k_0Ut$ and $\mathcal{U}'_N = \lambda u_{N+1}/U$, therefore, providing the original functions $u_n(t)$ through linear scaling. 
One can say that formulas (\ref{eqS2b_1})--(\ref{eqS2b_2}) introduce a hierarchy of intrinsic times and rescaled velocities associated to different scales of motion. 

System (\ref{eqSM}) in the inviscid unforced case ($\nu = 0$ and $f_n = 0$) is written for the dimensionless variables (\ref{eqS2}) as
	\begin{equation}
	\label{eqS3}
	\begin{array}{l}
	\displaystyle
	\frac{d\mathcal{U}_N}{d\tau} = 
	i\Big(k_{N+1}\mathcal{U}_{N+2}\,\mathcal{U}_{N+1}^*
	-\frac{1}{2}k_N\mathcal{U}_{N+1}\mathcal{U}_{N-1}^* \\[10pt]
	\displaystyle
	\qquad\qquad 
	+\,\frac{1}{2}k_{N-1}\mathcal{U}_{N-1}\mathcal{U}_{N-2}\Big)+\xi \,\mathcal{U}_N,
	\end{array}
	\end{equation}
where the factor $\xi = dT_m/dt$ is expressed using (\ref{eqS1}) and (\ref{eqS2}) by the series
	\begin{equation}
	\label{eqS4}
	\begin{array}{l}
	\displaystyle	
	\xi = \sum_{N < 0} k_N^3
	\left(2\pi_{N+1}-\frac{\pi_N}{2}
	-\frac{\pi_{N-1}}{4}\right),\\[17pt]
	\displaystyle
	\pi_N = \mathrm{Im}\left(\mathcal{U}_{N-1}^*\mathcal{U}_N^*\,\mathcal{U}_{N+1}\right).
	\end{array}
	\end{equation}
Note that the sum in (\ref{eqS4}) converges as a geometric progression with the dominant contribution from shells close to $N = 0$. To see this, one should write $\pi_N = k_{m}^3  T_m^3 \mathrm{Im} \left(u_{N+m-1}^*u_{N+m}^*u_{N+m+1}\right)$ and use the well-known scaling of third-order moment, $\langle \mathrm{Im}\,u_{n-1}^*u_n^*u_{n+1} \rangle \propto k_n^{-\zeta_3}$ with $\zeta_3 = 1$~\cite{l1998improved}. In particular, the sum in (\ref{eqS4}) can be considered in the inertial interval, neglecting contributions from the forcing scales. The new equation (\ref{eqS3}) is autonomous and contains quadratic and quartic nonlinearities. In the viscous and forced case, the corresponding extra terms in (\ref{eqS3}) receive time-dependent pre-factors depending on $T_m$. 

\textbf{Restored scale invariance for intrinsic variables.} Equations (\ref{eqS3})--(\ref{eqS4}) provide the equivalent formulation of the inviscid system, where the rescaled variables are introduced using the momentary temporal scale at a chosen reference shell $m$. The fact that these equations do not depend on $m$ signifies that the new system has the scaling symmetry related to an arbitrary choice of reference shell $m$. The generator of this symmetry corresponding to $m \mapsto m+1$ is given explicitly by (\ref{eqS2b_1}) with the inverse map given by (\ref{eqS2b_2}); this scale invariance can also checked by the direct substitution of (\ref{eqS2b_1})--(\ref{eqS2b_2}) into (\ref{eqS3})--(\ref{eqS4}). The obtained scaling relations (\ref{eqS2b_1})--(\ref{eqS2b_2}) are nonlinear both for temporal variable and shell velocities.  
Note that the symmetry transformations are local in scale, in the sense that they depend only on the variables at shells $0$ and $N$ or their nearest neighbors. 

The new symmetry is not necessarily broken, since intermittent fluctuations at each shell are absorbed into definition of the respective intrinsic time $\tau$. This observation also follows from the multiplicative form of scaling relations (\ref{eqS2b_1}); recall that multiplicative stochastic processes naturally accommodate  intermittency~\cite{friedrich1997description,touchette2009large}. We claim that this new scaling symmetry is restored in a statistical sense within the inertial range of developed turbulence. This statement is now being verified numerically by demonstrating the independence of statistical properties of variables $\mathcal{U}_N(\tau)$ on the choice of reference shell $m$. 

In the numerical analysis, we consider system (\ref{eqSM}) with constant forcing $f_1 = 1+i$, $f_2 = f_1/2$ and the total number of $40$ shells. The viscosity $\nu = 10^{-12}$ and parameters $k_0 = 1$, $F = 1$ are chosen such that $\mathrm{Re} = 10^{12}$ and the Kolmogorov dissipation scale corresponds to the shell $n \approx 30$. 
The initial condition is taken randomly from the statistically stationary state, which is then simulated with high accuracy in the interval $0 \le t \le 100$. In the analysis, we focus on the reference shells $m = 12,\ldots,21$ spanning three decades of scales at the center of inertial interval; this range is optimized to reduce the effects of insufficient statistics at lower scales and of intermittent dissipation at larger scales.

\begin{figure*}
\centering
\includegraphics[width=0.41\textwidth]{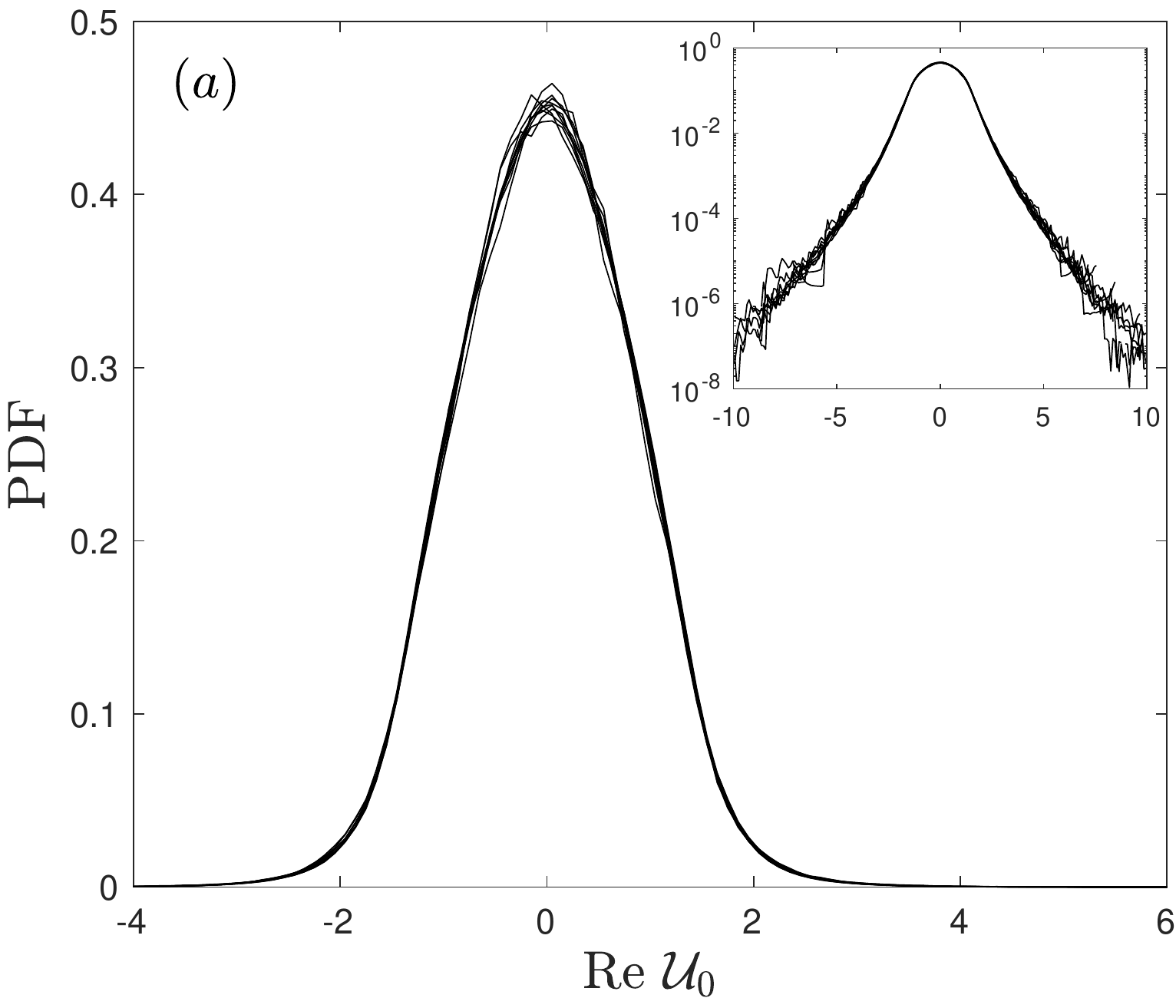}
\hspace{8mm}
\includegraphics[width=0.41\textwidth]{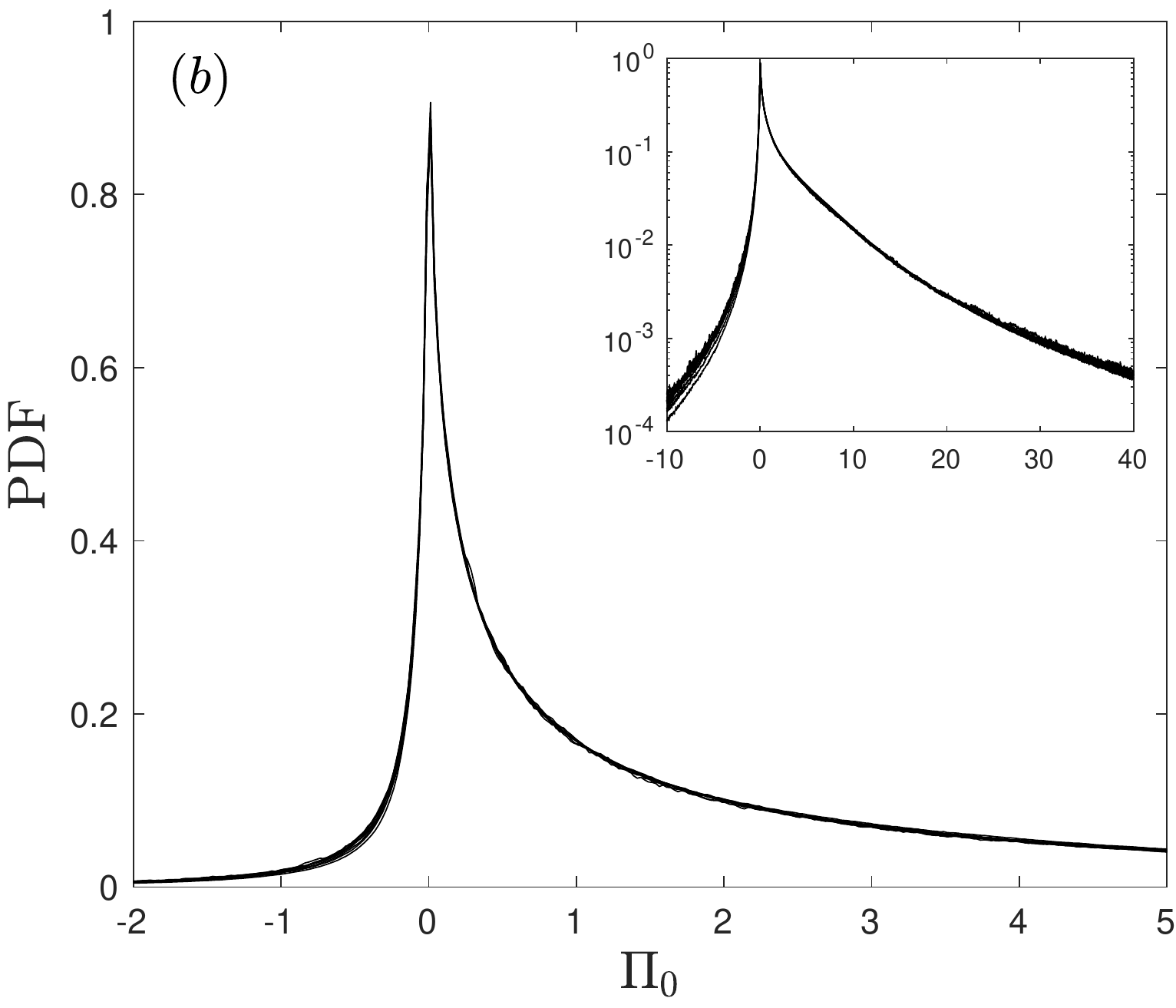}
\\
\vspace{4mm}
\includegraphics[width=0.41\textwidth]{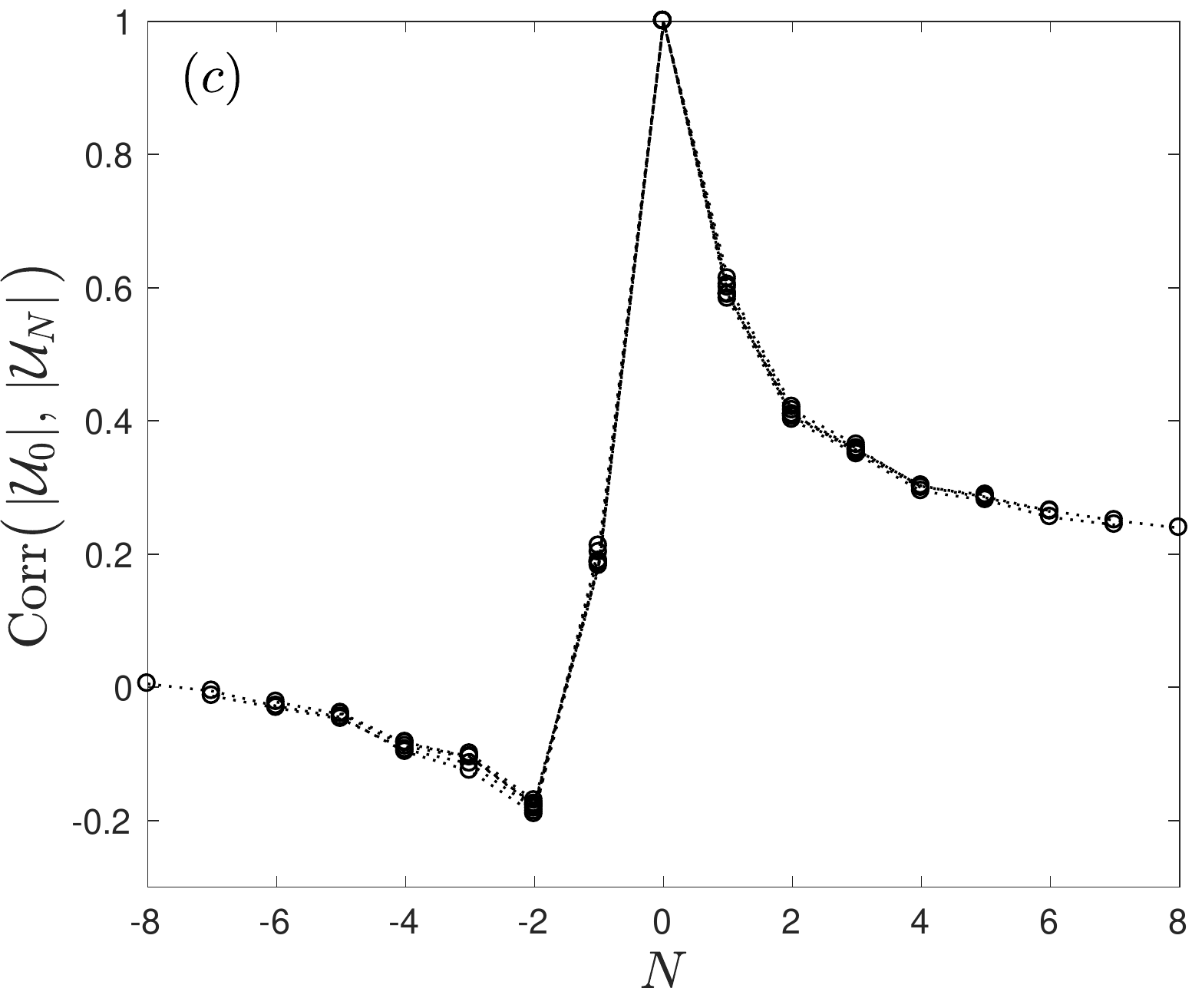}
\hspace{8mm}
\includegraphics[width=0.405\textwidth]{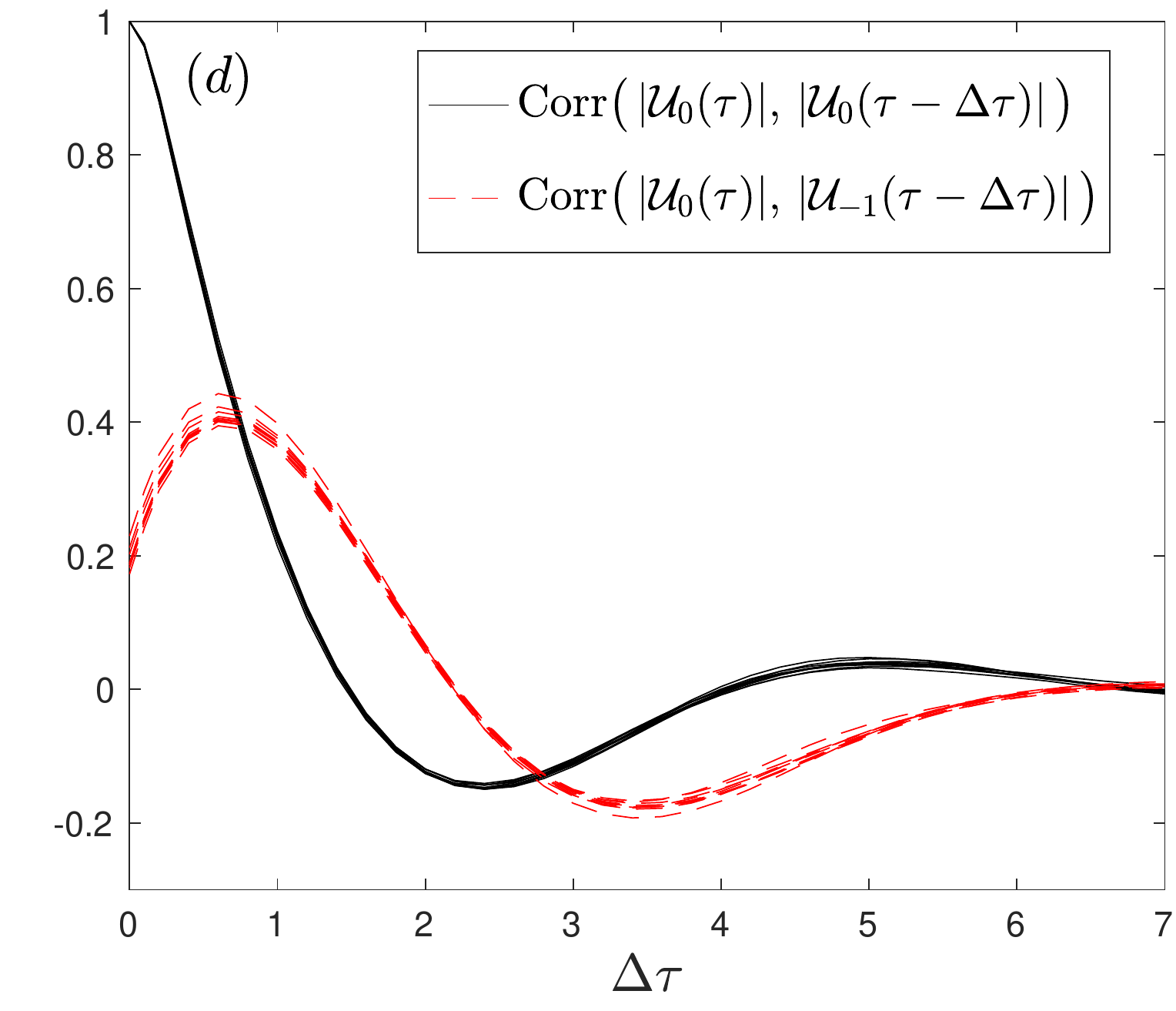}
\caption{Numerical demonstration of hidden scale invariance for intrinsic variables $\mathcal{U}_N(\tau)$. It is manifested as  independence of statistics on the reference scale (shell $m$) in the inertial range of developed turbulence. (a) Comparison of PDFs for $\mathrm{Re}\,\,\mathcal{U}_0(\tau)$, showing ten graphs for $m = 12,\ldots,21$. The inset repeats the same graphs in logarithmic vertical scale. (b) Analogous results for PDFs of dimensionless energy flux $\Pi_0(\tau)$ with $m = 12,\ldots,21$. (c) Correlation among scales: shown are correlation coefficients of $|\mathcal{U}_0(\tau)|$ and $|\mathcal{U}_N(\tau)|$ depending on $N$ for $12 \le N+m \le 21$ and $m = 13,\ldots,20$.  (d). Correlation in time: black curves show correlation coefficients between $|\mathcal{U}_0(\tau)|$ and $|\mathcal{U}_0(\tau-\Delta\tau)|$ taken at different fixed time delays $\Delta\tau$. Correlation across both scale and time: shown are correlation coefficients  between $|\mathcal{U}_0(\tau)|$ and $|\mathcal{U}_{-1}(\tau-\Delta\tau)|$. In both cases, $m = 12,\ldots,21$.}
\label{fig2}
\end{figure*}

We start by analyzing statistics of principal variables $\mathcal{U}_0(\tau)$ in Fig.~\ref{fig2}(a), where probability density functions (PDFs) of real parts are shown. The accurate collapse of all ten graphs for $m = 12,\ldots,21$ to a single curve brings the first strong evidence of the scale invariance. PDFs have exponential tails shown in the inset. Statistics of imaginary parts lead to similar results. The next Fig.~\ref{fig2}(b) presents PDFs for the cubic quantity 
$\Pi_0(\tau) = \pi_0+4\pi_1$ in the same range of shells, with the apparent scale invariance. Here $\Pi_0$ represents a dimensionless energy flux through the reference shell $m$~\cite{l1998improved,biferale2003shell}. PDFs feature positive mean values characteristic to direct energy cascade and resulting from  phase correlations~\cite{benzi1993intermittency,eyink2003gibbsian}; their tails are shown in the inset.

The last two figures examine correlations among different scales and times. Fig.~\ref{fig2}(c) provides the correlation coefficients between absolute values $|\mathcal{U}_0(\tau)|$ and $|\mathcal{U}_N(\tau)|$ for different $N$. The scale invariance implies that these correlation coefficients do not depend on the reference shell $m$, as the figure clearly suggests. Solid black curves in Fig.~\ref{fig2}(d) confirm the scale invariance for temporal correlations: they represent correlation coefficients between $|\mathcal{U}_0(\tau)|$ and $|\mathcal{U}_0(\tau-\Delta\tau)|$, i.e., at two values of the intrinsic time with the fixed delay $\Delta\tau$. Finally, red dashed curves in Fig.~\ref{fig2}(d) test the correlation simultaneously in scale and time by showing correlation coefficients between $|\mathcal{U}_0(\tau)|$ and $|\mathcal{U}_{-1}(\tau-\Delta\tau)|$, i.e., for the variables taken at different scales and different intrinsic times. All these tests confirm decisively that the proposed scale invariance is restored in a statistical sense. 

\textbf{Conclusion.} We established a hidden nonlinear scaling symmetry, which uncovers self-similarity of intermittent developed turbulence. This symmetry refers to inviscid equations reformulated for a hierarchy of intrinsic times at different scales of motion. The analysis is carried out for a shell model of turbulence and, therefore, the emerging self-similarity copes only with the temporal intermittency. However, what matters most is the clear evidence that intermittent dynamics can be rendered scale invariant with a proper definition of scaling relations. Experimental and numerical confirmation of universality for Kolmogorov multipliers~\cite{chen2003kolmogorov} indicates that analogous hidden symmetry may exist in the three-dimensional Navier--Stokes turbulence, where the spatial intermittency should be circumvented following similar guidelines. That is, using an additional hierarchy of scaling transformations in physical space with dilation rates controlled by intrinsic local scales. We notice here some parallel with conformal transformations in two-dimensional turbulence~\cite{bernard2006conformal}. Since the broken classical scale invariance is among major obstacles for describing turbulent flows~\cite{frisch1999turbulence}, the established form of self-similarity opens new opportunities for theoretical developments and applied modeling.

\bibliography{refs}

\end{document}